\documentclass[preprint,showpacs,preprintnumbers,amsmath,amssymb]{revtex4}
\usepackage{graphicx}
\usepackage{dcolumn}
\usepackage{bm}
\begin{document}
\newcommand{\be}{\begin{equation}}
\newcommand{\ee}{\end{equation}}
\newcommand{\bea}{\begin{eqnarray}}
\newcommand{\eea}{\end{eqnarray}}
\newcommand{\non}{\nonumber}
\title{Superluminal dispersion relations and the Unruh effect}
\author{Massimiliano Rinaldi}
\affiliation{ Dipartimento di Fisica, Universit\`a di Bologna and 
 I.N.F.N.\\
V. Irnerio 46, 40126 Bologna, Italy.\\
D\'epartement de Physique Th\'eorique, Universit\'e de Gen\`eve, \\
24, quai E. Ansermet 1211 Gen\`eve 4, Switzerland.} 
\date{\today}
\begin{abstract}
\noindent In the context of quantum gravity phenomenology, we study the Unruh effect in the presence of superluminal dispersion relations. In particular, we estimate the response function and the probability rate for an accelerated detector coupled to a massless scalar field, whose dispersion relation becomes essentially quadratic beyond a threshold momentum $k_p$. By means of a perturbative analysis, we show that superluminal dispersion induces a correction to the Planckian spectrum, which tends to vanish as $k_p$ increases.
\end{abstract}


\maketitle

\section{Introduction}

\noindent In recent years, the possibility that Lorentz invariance does not hold at very small scales has attracted increasing attention. In fact, the short-distance behavior of the field propagator is crucial, when extremely high energy phenomena are considered, and field modes probe distances of the order of the Planck length. Thus, the so-called trans-Planckian problem has been studied both in the context of cosmology and black hole physics (see, for example, \cite{cosm} and \cite{jacBH,bh}). A different approach to the trans-Planckian problem consists in introducing a minimal Lorentz-invariant length scale, which is the result of a duality symmetry of the path integral \cite{Pepe}. As we are interested in Lorentz-breaking theories, we do not consider this possibility here.

One way to implement the breaking of the Lorentz invariance is to consider modified dispersions, so that high energy modes propagate faster (or slower) than light. One motivation for these models is the analogy with super-fluids dynamics, where low energy (acoustic) excitations propagate like massless scalar fields on a curved background, with linear dispersion relation \cite{rivista}. However, at high frequency, the dispersion relation is no longer  linear (see, for example, \cite{landau}). On the gravity side, modified dispersion relations can be associated to the existence of a preferred frame, in such a way that general covariance is preserved \cite{jacobson}. In general, violations of the Lorentz symmetry are considered in several fundamental theories, such as loop quantum gravity or string theory (for a review, see \cite{matt}). 

Modifications of the dispersion relation can be seen as a phenomenological approach to quantum gravity, as far as they do not violate experimental bounds or well-established theoretical results. It is therefore important to examine such modifications in relation to various physical models. One example is the Unruh effect. It is well known that an inertial particle detector in flat space is not excited by field quanta when these are in the Minkowski vacuum state - a direct consequence of the  Poincar\'e-invariance of this state. On the contrary, when the detector follows an hyperbolic trajectory, it responds as it was in equilibrium with a thermal bath. In fact, the detector measures frequencies with respect to its proper time. Thus, the definition of positive frequency of an accelerated observer is not equivalent to that of an inertial observer  \cite{BirDav}. The natural question to ask is what happens to the accelerated detector,  when the dispersion relation is modified. Is the thermal character preserved? In particular, as the vacuum is no longer Poincar\'e-invariant, does the \emph{inertial} detector see field quanta?

In this paper, we examine these questions for a simple superluminal dispersion relation of the form $\omega_k=|\vec k|(1+|\vec k|^2/ k_p^2)^{1/2}$, where $k_p$ is of the order of the Planck energy. This model represents the simplest superluminal dispersion, which preserves rotational invariance in momentum-space. The term $|\vec k|^2/ k_p^2$ can also be seen as the first term of an expansion of a more complicate, and analytic, dispersion function $F(|\vec{k}|^2)$. In the final section, we briefly discussed also the dispersion model proposed by Unruh, which displays a constant frequency above a certain scale, see \cite{constdisp}.

We begin in the next section by  constructing the modified Wightman functions, which are necessary to evaluate the detector response. As exact solutions are not available, we expand perturbatively these functions in order to separate them in a Lorentz-invariant part plus a correction. To justify our approach, we stress that Unruh's effect is a low energy phenomenon, and this holds also when superluminal dispersion is considered. As the modified Wightman function decays very rapidly with $k$, the largest contribution to the corrections to the thermal spectrum comes from the regime $0<k<<k_p\,$.

In section 3, we evaluate the detector's response function and the probability rate of absorption. First, we look at the inertial case, and we check that the response vanishes identically for any superluminal dispersion. This a simple but crucial check. In fact, if the inertial detector registers field quanta, then one could have continuous particle creation in an adiabatic expanding Universe, and the existence of superluminal dispersion relations would be seriously questioned. Finally, we evaluate the response function for an accelerated detector, and find that superluminal dispersion yields deformations of the Planckian spectrum, which vanish as the threshold $k_p$ increases. The calculations are shown with some details in the two-dimensional case, and we only report the results for the 4-dimensional case, as they are obtained in a very similar way. 

In the last section, we discuss our results, and we show how they are consistent with the robustness of Hawking radiation against modifications of the dispersion relation.

\section{Modified Wightman's functions in Minkowski space}

\noindent In this section we compute the Wightman functions for a massless, minimally coupled scalar field $\phi(t,\vec{x})$, in the case when the relation dispersion has the form
\bea\label{MDR}
\omega_k=|\vec k|\sqrt{1+|\vec{k}|^2/k_p^2}\ .
\eea
An exact analytic form is hard to find. However, when the momentum is much smaller than the cut-off $k_p$, an approximate expression exists in the form of a correction to the Lorentz-invariant two-point function. In two dimensions, and in flat space, the positive  frequency Wightman functions are defined as the Fourier transform
\bea
G^{+}_2(x^{\mu},x'^{\mu})=\langle 0|\hat\phi(x^{\mu}),\hat\phi(x'^{\mu})|0\rangle =\int {d\vec{k}\over\sqrt{4\pi\omega_k}}\, e^{-i\omega_k(t-t')+i\vec{k}\cdot(\vec{x}-\vec{x}')}\ ,
\eea
which can be written as
\bea\label{twopoints}
G^{+}_2(t,x;t',x')=\int_{\Lambda}^{+\infty}{dk\over 2\pi\omega_k}\, e^{-i\omega_k\Delta t}\cos(k\Delta x)\ ,
\eea
where $\Delta t=t-t'$, $\Delta x=x-x'$, $\Lambda$ is an infra-red regulator, $|\vec{k}|=k$, and the usual ``$-i\epsilon$'' prescription is understood. In the Lorentz-invariant case (i.e. $\omega_k=k$), and for small $\Lambda$, the above integral yields
\bea\label{grel}
G^{+}_2(t,x;t',x')=-{1\over 4\pi}\ln |(\Delta t-i\epsilon)^2-\Delta x^2|+\rm{const}\ .
\eea
When $\omega_k$ has the form (\ref{MDR}), the integrand in Eq.\ (\ref{twopoints}) falls off rapidly as $k\rightarrow k_p$. Therefore, in the regime $k<<k_p\,$, we can expand the integrand for small $k/k_p\,$, so that the modified Wightman function can be written as
\bea\label{G2}
G^+_2=\int_{\Lambda}^{L} {dk\over 2\pi}\, H(\gamma, k)\ ,
\eea
where
\bea
H(\gamma,k)\simeq \cos(k\Delta x)\left[ {1\over k}\,e^{-ik\Delta t}-{k\over 2k_p^2}(1+ik\Delta t)\, e^{-ik\Delta t }\right]\ .
\eea
The first term corresponds to the Lorentz-invariant part of the propagator, hence we can extend the $k$-integration up to $L\rightarrow+\infty$, to find the expression (\ref{grel}), which we call $G_{0,2}^+$ . The second term is integrated over $k$ in the range $[0,L<<k_p]$, as it vanishes for $k\rightarrow 0$, and yields the correction
\bea\label{2g}
G^+_{k_p, 2}={1\over 4\pi k_p^2}\,\frac{(3\Delta t^4+6\Delta t^2\Delta x^2-\Delta x^4)}{(\Delta t^2-\Delta x^2)^3}+f_2(L)\ ,
\eea
where  $f_2$ is an oscillating function of the arbitrary parameter $L$. 
As a result, we can write the modified Wightman function as the sum of the relativistic expression and a first-order correction, according to $
G^+_2=G_{0, 2}^++G_{k_p, 2}^+\,$.

In four dimensions, very similar calculations give the result
\bea\label{corr4}
G^+_{k_p, 4}={1\over 4\pi^2k_p^2}\,\frac{(15\Delta t^4+10\Delta t^2\Delta x^2-\Delta x^4)}{(\Delta t^2-\Delta x^2)^4}+f_4(L)\ ,
\eea
where, again, $f_4$ is an oscillating function of  $k_p\,$. We observe that, due to the modified dispersion relation, the corrections to the Wightman functions are no longer a function of the Lorentz-invariant distance $(\Delta t^2-\Delta x^2)\,$.

\section{Particle detector}

\noindent In the standard treatment, a particle detector following a trajectory parameterised by the function $x^{\mu}(\tau)$, where $\tau$ is the detector's proper time, shows a transition amplitude to the first excited level given by the formula 
\bea
A\simeq g^2\int_{-\infty}^{+\infty}d\tau\, e^{\,i\Delta E\tau}\langle 1_{\bf{k}}|\hat \phi(x)|0\rangle\ .
\eea
The parameter $g$ represents the coupling between the detector and the field, and the above expression is valid when $g$ is sufficiently small \cite{BirDav}. Also, $\Delta E=E-E_0>0$ is the energy gap between the first excited state and the ground state of the detector, and $|1_{\bf{k}}\rangle$ is the one-particle state. With the help of the usual commutation rules, and in $n$ dimensions, one finds
\bea
\langle 1_{\bf{k}}|\hat \phi(x)|0\rangle=\left (2\omega_k(2\pi)^{n-1}\right)^{-1/2}\, e^{-i\vec{k}\cdot\vec{x}+i\omega_k t}\ ,
\eea
and the amplitude is given by
\bea
A\simeq {g^2\over \left(2\omega_k(2\pi)^{n-1}\right)^{1/2}}\int_{-\infty}^{+\infty}d\tau\, e^{i\Delta E \tau}\, e^{-i\vec{k}\cdot\vec{x}+i\omega_k t}\ .
\eea
This integral vanishes identically for an inertial detector in the Lorentz-invariant case. As discussed in the introduction, it is important to check that this property holds also when the dispersion is modified. So, let $ \vec{x}=\vec {x}_0+\vec{v}(1-v^2)^{-1/2}\tau\equiv \vec {x}_0+\gamma\tau\vec v$, where $|\vec v|<1$ and $0<\gamma<1$. It follows that the $\tau$-integral is proportional to a delta  function, and
\bea
A\propto \delta(\Delta E+\gamma(\omega_k-\vec k\cdot\vec v))\ .
\eea
In the Lorentz-invariant case, $\omega_k=|\vec{k}|>|\vec k||\vec v|>\vec k\cdot\vec v$, hence the above function vanishes identically \cite{BirDav}. It is easy to see that whenever the dispersion has a form such that $\omega_k>|\vec k|$, the amplitude vanishes  \footnote{This simple argument also shows that, for subluminal dispersion, the amplitude might not be zero. On one hand,  this could lead to instabilities of the model \cite{rashidi}. On the other hand, this can be interpreted instead as a sort of \v{C}erenkov effect, see the discussion in \cite{matt}}. Therefore, the amplitude is zero also for the dispersion (\ref{MDR}).

We now consider a non-inertial detector. By squaring the amplitude and by integrating over all $k$, one finds the detector response function, given by \cite{BirDav}
\bea\label{response}
F=g^2\int_0^{T}d\tau\int_0^{T}d\tau'\, e^{-i\Delta E\Delta\tau}G^+(x^{\mu}(\tau),x^{\mu}(\tau'))\ .
\eea
In the Lorentz-invariant case, $G^+$ is a function of $\Delta\tau$ only. Hence, differentiation with respect to $T$ yields the probability rate of absorption per unit proper time, namely
\bea\label{rate}
R_-=g^2\int_{-\infty}^{+\infty}d(\Delta\tau)\, e^{-i\Delta E\Delta\tau}\, G^+(\Delta\tau-i\epsilon)\ .
\eea

As we have seen in the previous section, when the dispersion is modified, the corrections to the Wightman's function are no longer a function of the Lorentz-invariant distance $\Delta t^2-\Delta x^2$. Therefore, when we choose the hyperbolic trajectory 
\bea\label{traj}
x=(t^2+\alpha^{-2})^{1/2}\ ,\qquad t= \sinh (\tau\alpha)/\alpha\ .
\eea
where $\alpha$ is the acceleration, the Wightman's function is not a function of $\Delta \tau$ only, thus we need to compute explicitly the response (\ref{response}), before finding the probability rate \footnote{In four dimensions, the hyperbolic trajectory is also defined by  $z=y=0$.}. In the Lorentz-invariant case instead, the probability rate of absorption can be found directly from Eq. (\ref{rate}). In two and four dimensions, one finds respectively \cite{BirDav}
\bea\label{relrate2}
R_-^{(2)}={g^2\over \Delta E}\left(e^{2\pi\Delta E/\alpha}-1\right)^{-1}\ ,
\eea
and
\bea\label{relrate4}
R_-^{(4)}={g^2\Delta E\over 2\pi}\left(e^{2\pi\Delta E/\alpha}-1\right)^{-1}\ .
\eea

Let us look at the 2-dimensional case, and consider the correction (\ref{2g}). When the detector follows the hyperbolic trajectory (\ref{traj}), the response function associated to the correction becomes
\bea
F_2={g^2\over 4\pi k_p^2}\int_0^{T}d\tau\int_{0}^{T}d\tau'{(8C^4-4C^2-1)\over S^2}\,e^{-i\Delta E(\tau-\tau')}\ ,
\eea
where
\bea
C=\cosh\left({\alpha(\tau+\tau')\over 2}\right)\ ,\qquad S={2\over \alpha}\sinh\left({\alpha(\tau-\tau')\over 2}\right)\ .
\eea
In these calculations, we have ignored the boundary terms coming from $f_2(L)$, as they yield infinite oscillating functions. We now change variables
\bea
r+s=2\tau\ ,\qquad r-s=2\tau'\ ,
\eea
and we extend the integration over $s$ to  $\pm\infty$,  as the main contribution to the integral comes from $s={\cal O }(\Delta E^{-1})$, \cite{BirDav}. Thus, we find
\bea
F_2=-{\alpha g^2\over 16\pi k_p^2}f_2(\alpha T)\int_{-\infty}^{+\infty}ds\, {e^{-i\Delta E s}\over \sinh^2({\alpha s\over 2}-i\epsilon)}\ ,
\eea
where $-i\epsilon$ is added to cope with the pole, and where we defined
\bea
f_2(\alpha T)=\cosh\left({\alpha T\over 2}\right)\sinh\left({\alpha T\over 2}\right)\left[2\cosh^2 \left({\alpha T\over 2}\right)+1\right]\ .
\eea
By integrating in $s$, we find
\bea
F_2={\Delta Eg^2\over 2 k_p^2}{f_2(\alpha T)\over\alpha}\left(e^{2\pi\Delta E/\alpha}-1\right)^{-1}\ .
\eea
To compute the probability rate per unit proper time, we divide by $T$, and choose $\alpha T$ small enough so that 
\bea\label{approx}
{f_2(\alpha T)\over \alpha T}={3\over 2}+{\cal O}(\alpha^2 T^2)\ .
\eea
Thus, by adding also  the Lorentz-invariant rate of absorption (\ref{relrate2}), we find the total rate 
\bea
R_-^{(2)}\simeq {g^2\over \Delta E}\left(1+{3\Delta E^2\over 4 k_p^2}\right)\left(e^{2\pi\Delta E/\alpha}-1\right)^{-1}\ .
\eea

In four dimensions, we analyze the correction (\ref{corr4}). Analogous calculations leads to
\bea
F_4=-{g^2\over 8\pi k_p^2}f_4(\alpha T)\int_{-\infty}^{+\infty}ds{e^{-i\Delta M s}\over \sinh^4\left({\alpha s\over 2}-i\epsilon\right)}\ ,
\eea
where now
\bea
f_4(\alpha T)=4T+{2\over\alpha}\sinh\left({\alpha T\over 2}\right)\cosh\left({\alpha T\over 2}\right)\left[ 6\cosh^2\left({\alpha T\over 2}\right)+5 \right]\ .
\eea
By integrating by parts, and discarding infinite oscillating terms, we find
\bea\non
I&=&\int_{-\infty}^{+\infty}ds{e^{-i\Delta M s}\over \sinh^4\left({\alpha s\over 2}-i\epsilon\right)}={4i\Delta M\over 3\alpha}\int_{-\infty}^{+\infty}ds{e^{-i\Delta M s}\cosh\left({\alpha s\over 2}\right)\over \sinh\left({\alpha s\over 2}-i\epsilon\right)}+\\\non\\
&+&{2\Delta M^2\over 3\alpha^2}\int_{-\infty}^{+\infty}ds{e^{-i\Delta M s}\over \sinh^2\left({\alpha s\over 2}-i\epsilon\right)}\ ,
\eea
which is a sum of Planckian-type integrals, and gives
\bea
I=-{16\pi \Delta E\over 3\alpha^2}\left(1+{\Delta E^2\over \alpha^2}\right)\left(e^{2\pi\Delta E/\alpha}-1\right)^{-1}\ .
\eea
To compute the rate of absorption, we divide again $F_4$ by $T$ and take $\alpha T$ small, so that
\bea\label{fexp}
f_4(\alpha T)\simeq 15+{\cal O}(\alpha^2 T^2)\ .
\eea
The final result, which includes the Lorentz-invariant contribution (\ref{relrate4}), reads
\bea\label{corr4d}
R_-={g^2\Delta E\over 2\pi}\left[1+{5\pi\alpha^2\over 4k_p^2}\left(1+{\Delta E^2\over \alpha^2}\right)\right]\left(e^{2\pi\Delta E/\alpha}-1\right)^{-1}\ .
\eea

\section{Discussion}

\noindent In the previous sections, we estimated the modifications  induced by superluminal dispersion relations to the Planckian spectrum detected by an accelerated observer. As an exact calculation is hard to perform, we approximated the integrals over the momentum by integrating up to a cut-off $L$.  The dispersion relation studied in this work become essentially quadratic as $k$ is of the order of $k_p>>L$. However, the modified Wightman's function has a peak for $k<<L$, and the biggest contribution to the probability rate comes from the low-momentum regime, just as in the Lorentz-invariant case. Thus, Unruh's effect remains a low energy phenomenon also when superluminal dispersion is present. The corrections that we have found to the Planckian spectrum are proportional to $1/k_p^2$. Hence, when $k_p$ is large, these corrections become very small. For example, in four dimensions, the correction (\ref{corr4d}) essentially depends on the ratio $\alpha^2/ k_p^2$. If $\alpha$ is of the order of $\Delta E$ (few eV), so that the Planckian factor is not trivial, and $k_p$ is of the order of the Planck energy $\sim 10^{19}$ GeV, we see that the corrections become negligible.

This result is not surprising, in the light of the well-known robustness of Hawking radiation with modified dispersion relations. It is known that the near-horizon region of a black hole can be mapped to a Rindler space. Therefore, a detector placed just outside the horizon behaves as a static detector in Rindler space. In turn, this is equivalent to a detector in Minkowski space, moving along the hyperbolic trajectory defined by Eq.\ (\ref{traj}) \cite{UnruhD}. Thus, the thermal bath detected by the accelerated observer is essentially equivalent to the Hawking emission near the horizon. When the dispersion relation is modified, the latter shows a negligible deviation from thermality. The proof of this result, presented in \cite{bh}, requires that the surface gravity $\kappa$ of the black hole and the distance $x$ from the horizon satisfy the constraints $\kappa \ll k_{\rm cutoff}$ and $|x|k_{\rm cutoff} \gg1$, where $ k_{\rm cutoff}$ is the scale at which the dispersion relation departs from linearity. As, in the near-horizon region, the acceleration of a static observer is given by $\alpha\simeq\sqrt{\kappa/x}$, these constraints imply $\alpha/ k_{\rm cutoff}\ll 1$. In our setup,  $k_{\rm cutoff}$ is the same as $k_p$, and $\alpha$ is the acceleration of the detector moving in flat space along the trajectory (\ref{traj}). Therefore the comparison between the radiation emitted near the horizon and the one detected by an observer in flat space only makes sense when the acceleration of the latter is such that $\alpha/k_p\ll 1$. In this regime, on one hand we find that the deviation from thermality in Eq.\ (\ref{corr4d}) is in fact negligible, and this is fully consistent with the robustness of the Hawking radiation. On the other hand, our results also enforce the similarity between the Unruh effect and the Hawking evaporation. These effects are known to be related in the case of linear dispersion relations. Our calculations show that this relationship holds also in the presence of modified dispersion relations.

We finally comment on the dispersion relations studied in \cite{constdisp}. In these models, the frequency $\omega$ becomes constant above a certain scale, therefore the energy is bounded from above even in the limit of infinite momenta. This possibility must be distinguished from the models mentioned in the introduction, which assume a maximum momentum, corresponding to a minimal, covariant length \cite{Pepe}. In the context of our calculations, we could model the dispersion of \cite{constdisp} as being linear or superluminal for $k<L$ and constant for $k>L$, where $L$ was defined in section 2. Thus, the Wightman function (\ref{twopoints}) would be modified and, in the range $[L,\infty]$, we would have $\omega_k=\,\,$const. As a result, the integral is no longer well-defined, as the exponential damping factor becomes constant. On a physical ground, one can interpret this infinite oscillating function in the same way as the function $f_2(L)$ defined in Eq.\ (\ref{2g}), in the sense that the net contribution to the Unruh effect is vanishing on the average. This interpretation would be consistent with the relationship between the Unruh effect and the Hawking evaporation, as the constant $\omega$ at high momenta does not alter the spectrum of the radiation \cite{constdisp}. Finally, in the case of a dispersion, which is subluminal for $k<L$ and linear for $k>L$, one might encounter the stability problem mentioned above, and discussed thoroughly in \cite{rashidi,matt}.

\acknowledgments
\noindent I wish to thank  R.\ Balbinot,  J.\ Navarro-Salas, A.\ Fabbri, P.\ Anderson, R.\ Parentani, and T.\ Jacobson for useful discussions.

\end{document}